\documentclass{PoS}

\title{Effective scalar fields in Yang-Mills theories}

\ShortTitle{Effective scalar fields }

\author{Henri Verschelde$^\dagger$, \speaker{Valentin I. Zakharov}$^\ddagger$\\
	$^\dagger$Ghent University, Department of Physics and Astronomy,\\
        \hspace{0.15cm}Krijgslaan 281 - S9, 9000 Ghent, Belgium.\\
         $^\ddagger$Institute for Theoretical and Experimental Physics,\\
        \hspace{0.25cm}B. Cheremushkinskaya 25, Moscow, 117218, Russia.\\
        \hspace{0.15cm}Max-Planck-Institut f\"ur Physik,\\
        \hspace{0.25cm}F\"ohringer Ring 6, 80805 M\"unchen, Germany.\\
         \hspace{0.15cm}E-mail: \email{henri.verschelde@ugent.be}, \email{xxz@mppmu.mpg.de}}

\abstract{ Scalar fields play a crucial role in the Standard model. On the other hand,
in the weak-coupling regime
there is an unsolved problem of the quadratic divergence of scalar masses. Thus, it is natural
to turn to composite, or effective scalar fields in the strong-coupling regime. Lattice 
simulations provide information on the ``actually existing" scalar fields in Yang-Mills theories.
On the continuum-theory side, dual models predict existence of various low-dimension
vacuum defects, including probably scalar particles. We are looking for correspondence between 
the two frameworks, of lattice phenomenology and dual models, and discuss possible applications
to the theory of the Yang-Mills plasma in the deconfining phase. It is not ruled out that the 
effective scalars play important role in the plasma dynamics.  }

\FullConference{The many faces of QCD\\
		November 1-5, 2010\\
		Gent Belgium}

\begin{document}

\section{Introduction}
The title of the conference invites us to talk about various facets of QCD.
One of the most interesting and, nevertheless, rarely covered aspects
of QCD is the story about effective scalar, or Higgs-like particles. We are sure about the fundamental
Lagrangian of QCD (which we actually reduce, for our purposes, to its pure gluonic
 part), and there are no fundamental scalar fields
 in this Lagrangian. Unlike the theory of weak
 interactions, where one does introduce Higgs fields. In the case of weak coupling,
 however, there is the problem of the quadratic divergence in the mass,
 or in the vacuum expectation value of the Higgs field, $<|H|^2>\sim \Lambda_{UV}^2$.
 The authors of the talk belong to the theorists who consider this divergence
 a fatal failure of the theory. What we actually need to avoid in the hierarchy problem is
\begin{equation}\label{higgs}
<|H|^2>~\sim~\Lambda_{IR}^2
\end{equation}
where $\Lambda_{IR}$ is the mass scale associated, probably, with new strong interactions,
and the Higgs field would be then a composite particle.

In terms of QCD the question is then, whether there 
exist effective scalar degrees of freedom such that   (\ref{higgs}) is satisfied 
with $\Lambda_{IR}=\Lambda_{QCD}$.
   The path   to
 scalars  is provided by  considerations of the   confinement mechanism
 in Yang-Mills theories. In Abelian cases
 \cite{polyakov} the confinement
 of electric charges is due to
condensation of a magnetically charged scalar field.
By analogy, one is tempted to assume that in the non-Abelian
case the  tension of the string connecting  heavy quarks is also of the order
\begin{equation}\label{guess}
\sigma_{Q\bar{Q}}~\sim~<\Phi_M>^2~\sim~\Lambda_{QCD}^2~~~(expectation)
\end{equation}
where $\Phi_M$ is a magnetically charged composite field.
Although the guess (\ref{guess})
looks appealing, there are many questions left. What
is the precise meaning
of the magnetic degree of freedom, how is magnetic charge is defined,
which symmetry is spontaneously broken by the vacuum expectation value
$<\Phi_M>\neq 0$, and so on.
 
Magnetic degrees of freedom responsible for the 
confinement
have been searched for on the lattice,
for review   and further
references see, e.g., \cite{greensite}, and there exists extensive lattice phenomenology.
The interpretation of the lattice data, however, is not straightforward.
First, particles and strings on the lattice are to be described in terms of 
the quantum geometry
which deals with lines and surfaces fluctuating on the size of the lattice
spacing $a$, $a\to 0$ in the continuum limit,
see, e.g., \cite{polyakov3}. Moreover, there are also specific 
lattice-related constructions which have no direct analogy in the continuum limit.
Finally, every lattice configuration is to be understood as a measurement of
the Yang-Mills fields with resolution of the lattice spacing $a$. While
physical  matrix elements are immune to the details of the
measuring procedure, some other observations, like
localization of a fermionic zero mode might depend on the measuring procedure \cite{resolution},
as is common in quantum mechanics. Thus, the translation from the
 lattice to the continuum-theory language  is a long process and  we refer the reader to   \cite{theory} and references therein for further details.
 
 Confinement is a non-perturbative phenomenon and infrared physics is involved.
 In the field theoretic language the only known non-perturbative configuration  
 is the instanton.  The dual formulations, completing the Y-M theories in the infrared,
 involve extra dimensions
 and are much richer in topologically stable excitations, for particular examples
 and references see \cite{gorsky2,zhitnitsky}. 
  
  The talk is, essentially, in three parts. First, we review the lattice data on
  the scalars. Then, we 
 discuss effective scalar fields within the dual models. Finally, we are 
 looking for applications to the  the quark-gluon plasma,
 (for review of the plasma properties see, e.g., \cite {teaney}).
\section{Particles and strings on the lattice}
\subsection{Lattice monopoles, or scalars}
Monopoles are identified as certain trajectories  in the 4d volume of the lattice
\footnote{The name of ``monopoles" should not be taken too literally
and is used mostly for the historic reasons.}.
The properties of the trajectories can be  related to
field theoretic quantities  within the polymer approach to field theory. 
One of the simplest things to measure is the total length of the
monopole trajectories.
It turns out that the total length is described by a two-term expression
\cite{monopoles}:
\begin{equation}\label{monopoly}
(L_{tot})_{monopoles}~=~\Big(c_1\Lambda_{QCD}^3+c_2\Lambda_{QCD}^2a^{-1}\Big)V_{tot}~~,
\end{equation}
where $V_{tot}$ is the total volume of the lattice and the origin of this factor is
trivial, $c_{1,2}$ are  fitting parameters, for details and further
references see Ref. \cite{monopoles}. Note that the term proportional to $a^{-1}$ dominates
in the continuum limit. Nevertheless, we keep both terms and for a good reason. Namely,
the two terms in (\ref{monopoly}) refer to two different types of trajectories.
The first term refers to an infinite cluster, while the second term
corresponds to finite, or short clusters.
In field theoretic terms, an infinite cluster is  a condensate and short clusters are
quantum fluctuations.
It is worth emphasizing that the emergence of the parameter $\Lambda_{QCD}$
is a highly non-trivial observation since $\Lambda_{QCD}$
is related to the lattice spacing by means of the two-loop $\beta$-function.
The emergence of $\Lambda_{QCD}$ in numerical simulations provides a strong support
to the idea that lattice monopoles correspond indeed to physical objects.
\subsection{Very dilute monopole condensate }

Eq. (\ref{monopoly}) can be considered as the main message from the lattice to the
continuum theory. Indeed,  for  an elementary scalar one   would have
  \begin{equation}\label{elementary}
 L_{elem}~\sim~a^{-3}V_{tot},~~~~~<|\Phi_{elem}|^2>~\sim~a^{-2}
 \end{equation}
and reproduce the standard quadratic divergence which plagues the elementary-scalar
theories. In contrast to this, Eq. (\ref{monopoly}) implies for the vacuum
expectation of the lattice scalar:
 \begin{equation}\label{amusing}
 <|\Phi_M|^2>~\sim~\lim_{a\to 0}(a\cdot L_{tot})~\sim~c_2\Lambda_{QCD}^2~.
 \end{equation}
 This is an amusing result
 since (\ref{amusing}) is exactly what we would like to have
 for a composite Higgs fields, compare (\ref{higgs}). However, there are surprises left as we will
 immediately see.
 
  Indeed,
 for the classical part of the field, or the monopole condensate application
 of quantum geometry brings an unexpected result \cite{chernodub2002}:
 \begin{equation}\label{amazing}
 <\Phi_M>^2~\sim~(a\cdot \Lambda_{QCD})\Lambda_{QCD}^2~~,
 \end{equation}
 so that the classical part vanishes in the continuum limit $a\to 0$.
 The lattice result (\ref{amazing}) is in sharp contrast with the expectation
 (\ref{guess}).
Thus, Eq. (\ref{amazing}) suggests that confinement
 is ensured by a condensate which is vanishing in the continuum limit!
 Nevertheless, at any finite value of the lattice spacing $a$
 the monopole models do explain the string tension $\sigma_{Q\bar{Q}}$
 in terms of   the monopole condensate.
   The price is that
the models  assume long-range forces, or massless
 exchange in the Yang-Mills vacuum. Anyhow, we have to look for a substitution
 for $<\Phi_M>$ in Eq. (\ref{guess}). Probably, it is the string density which takes
 over $<\Phi_M>^2$, see below.
\subsection{Magnetic  surfaces, or strings}
Probably, the most unexpected observation concerning monopoles is that the
monopole trajectories do not percolate in fact through
the whole 4d volume $V_{tot}$ but rather lie on 2d surfaces which are defined
independently. Two-dimensional surfaces correspond to strings and one can say,
therefore, that the lattice monopoles live on strings.
The absence of the ultraviolet divergence in Eq. (\ref{amusing})
 is just a manifestation of this alignment of the monopole trajectories with the
 vortices.

 In more detail, the total area of the surfaces  is empirically given
by  \cite{branes,greensite}:
 \begin{equation}\label{physical}
(Area)_{tot}~\approx~c_3\Lambda_{QCD}^2V_{tot}~\equiv~\rho_{strings}V_{tot}.
\end{equation}
The surfaces are closed in the vacuum.
The monopoles, in turn,
 are closed trajectories lying on the surfaces.
Small clusters of monopole trajectories cover the surfaces densely
while the infinite cluster, still belonging to the surfaces, is dilute
and occupies a vanishing part of the total area (\ref{physical}).
In case of the surfaces one can also distinguish clusters, finite
or an infinite. However, unlike the monopole case,
  the total area (\ref{physical})
is dominated by a single, percolating cluster of the vortices (or strings).

From the fact that the area (\ref{physical}) scales in physical units,
see (\ref{physical}), one would conclude that the string tension is of
order
\begin{equation}\label{simplification}
T_{magmetic ~strings}~\sim~\Lambda_{QCD}^2~.
\end{equation}
In the continuum limit, the
lattice
strings are identified as magnetic strings
\cite{theory,gorsky2} since  they are closed in the vacuum and can be open on
an external
't Hooft line.


 \subsection{Deconfinement phase transition}
 At the deconfinement phase transition the monopole trajectories and magnetic 
 surfaces become time oriented, for review and references see \cite{greensite}.
 Namely, for the monopole trajectories one can introduce an asymmetry parameter
 \begin{equation}
 \nu~=~{L_{\tau}-1/3 L_{x+y+z,}\over L_{\tau}+1/3 L_{x+y+z,}}~~,
 \end{equation}
 where, for example,  $L_{\tau}=N_{\tau}\cdot a$  and $N_{\tau}$ is the number
 of links belonging to the percolating monopole trajectories and looking in
 the (Euclidean) time direction. A similar parameter can be introduced for 
 magnetic strings, in terms of the total areas occupied by 
 plaquettes belonging to the surfaces and oriented in various ways.
 
 At low temperatures the asymmetry $\nu=0$. At temperatures close to the 
 temperature of the phase transition $T_c$ there appears a non-zero
 $\nu$. And at about $T\sim 1.5 T_c$ the time-oriented links dominate.
 
 \section{Stringy models for Yang-Mills theories  }
 \subsection{Geometry of extra dimensions}
 
 A dual, or stringy description of Yang-Mills theories has not been 
 yet constructed. However, there exists a model \cite{witten} 
 which belongs to the same universality class as the large-$N_c$
 Yang-Mills theories in the far infrared.

The corresponding geometry   involves ten dimensions,
where the number of ten is fixed by the requirement to have a consistent
string theory
at short distances. For our purposes, we concentrate on the standard four dimensions, fifth dimension,
denoted here as $u$ and one
  extra compact coordinate, $x_4$. The fifth, mass scale-conjugated dimension is common to all dual models.  The limit $u\to \infty$ means short distances in our world while
    finite values of $u$ correspond
  to   poorer resolution in our world, or moving towards the infrared.
  The $x_4$ coordinate  is specific for the universality class  of
  the Yang-Mills theories \cite{witten}.
At temperature $T=0$ the metric is:
\begin{eqnarray}\label{sakai}
ds^2=\Big({u\over R}\Big)^{3/2}(-dt^2+\delta_{ij}dx^idx^j+f(u)dx_4^2)+
\Big({u\over R}\Big)^{3/2}({du^2\over f(u)}+u^2d\Omega_4^2)\\ \nonumber
~~~~~f(u)~=~1-\big({u_{\Lambda}\over u}\big)^3,~~~~x_4~\sim~x_4+\beta_4,~~~~\beta_4={4\pi\over 3}\Big({R^3\over u_{\Lambda}}\Big),
\end{eqnarray}
where $R$ is a constant related to the Y-M coupling constant.
Note the existence of a horizon at $u=u_{\Lambda}$ which is
crucial to ensure confinement. Moreover, the $x_4$ coordinate is related to the topological charge.
Namely, if a defect is wrapped once over the $x_4$ circle, it has a unit topological charge.
Wrapping in the opposite direction brings a minus sign for the topological charge.
Extra compact dimensions, associated with the unit sphere $\Omega_4$
are relevant to baryons and do not concern us here.

 At finite Euclidean temperatures,
  there are two compact
coordinates, $x_4$ and the Euclidean time, $\tau$.
The deconfinement phase transition is identified as the Hawking-Page transition
which interchanges the geometries in the $(x_4,u)$ coordinates and in
the $(\tau,u)$ coordinates \cite{aharony}.
Namely,
 at low temperatures the geometry in the $(u,x_4)$ coordinates is cigar-shaped 
 since the radius  of the $x_4$-coordinate
tends to zero at the horizon:
\begin{equation}\label{geometry1}
R_{\tau}(u)~=~{1\over 2\pi T}~,~~R_{x_4}(u_{\Lambda})~=~0~~;T<T_c~.
\end{equation}
On the other hand, the radius of
the Euclidean time direction is independent of $u$.  

At temperatures above the phase transition the geometry of the two
compact coordinates is interchanged so that:
\begin{equation}\label{geometry2}
R_{\tau}(u_T)~=~0~,~~R_{x_4}(u)~=~const~~;T>T_c~,
\end{equation}
where $u_T$ is the position of the temperature-related horizon.
The phase transition takes place at the  temperature where the two radii,
 namely, $R_{\tau}\equiv1/2\pi T$ and the value of $ \beta_4/2\pi$,
become equal to each other.

\subsection{Low-dimensional defects}

What kind of physics could the model (\ref{sakai}) describe?
At short distances, or at $u\to \infty$, the model has five dimensions,
instead of four and is apparently not relevant. 
The model can be valid only at distances $x\gg \beta_4~\sim~\Lambda_{QCD}^{-1} $,
or at values of $u$ close to the horizon. 

Thus, at $T>T_c$ we are left with {\it hydrodynamics} of the Yang-Mills plasma
as potential application of the model (\ref{sakai}). At temperature $T=0$
one could apply the model to the vacuum which is - in our context -
nothing else but the theory of low-dimensional defects, like monopoles
and magnetic strings. Moreover, in all the cases, it is more appropriate
to talk about the non-perturbative components (say, of the same plasma)
since the perturbative contributions come from short distances
and cannot be described within the model (\ref{sakai}).

Concerning the notion of defects, the  Polyakov line \cite{polyakov1}
can be considered as a prototype of such defects. It is defined in the Euclidean
 space  as
 \begin{equation}\label{polyakov}
 L({\bf x})~\equiv~{1\over N_c}~Tr~ P \exp\Big(-i\int_0^{1/T}A_0({\bf x},\tau)~d\tau\Big) ~~,
 \end{equation}
 where $\tau$ is the Euclidean (periodic) time, $A_0$ is the gluon field.
 Note that the Polyakov line depends only on the spatial coordinates
 since one integrates over the time direction along the line. From
 the point of view of the 3d theory the Polyakov line (\ref{polyakov})
 is  a point-like defect. It was argued first in Ref. \cite{polyakov1}
 that actually this defect might represent a dynamical degree of freedom
 at $T>T_c$ which is to be added to the Hamiltonian of the Yang-Mills theories
 by hand.
 Later, there appeared
   many models which try to make this picture more precise,
 see \cite{pisarski} and references therein.

In the models, like (\ref{sakai}) there are many more defects. Indeed,
according to the theory there exist D0, D2, D4 branes , for details and incomplete list
see, in particular, \cite{zhitnitsky}. According to the general rules, the probability
to find such defects is exponentially small in the limit of large $N_c$,
$$S_{defect}~\sim~N_c\cdot T\cdot (Volume)_{defect},$$ where $T$ is a generalized tension 
and $(Volume)_{defect}$ is the volume occupied by the defect. Such defects are not significant dynamically. However, there are cases
when the volume of the defect vanishes,
\begin{equation}\label{volume}
(Volume)_{defect}~=~0,~~~S_{defect}=0 
\end{equation}
in the classical approximation.
Such defects can become dynamical and, in particular, might correspond to the percolating
monopoles or magnetic strings. 

\subsection{Magnetic strings in dual models}

Note that all the branes wrapped around the $x_4$ compact dimension at temperature $T=0$
satisfy the condition (\ref{volume}) and can become dynamical in the infrared limit.
In particular,   consider D2 branes which are
wrapped around the $x_4$ coordinates
and extended in two other directions of the Euclidean 4d space. Then at zero temperature
the action associated with such defects is vanishing
in the infrared:
\begin{equation}
S_{D2~branes}(u)~\sim~R_{x_4}(u)\cdot (Area)_{4d}~,
\end{equation}
where $(Area)_{4d}$ is the area swept by the brane in the Euclidean 4d.
Because of (\ref{geometry1}) the brane action vanishes for the branes living on the horizon,
$u\to u_{\Lambda}$.

Let us interpret this observation. First of all, ``living
on the horizon " means large distances, or the infrared limit.
In field theoretic language, for example, instantons have action vanishing
in the infrared. In other words, they are not suppressed by the action and
populate the vacuum. Now, a similar picture holds for the 2D branes
wrapped around the $x_4$ circle. Thus, in 4d these branes could look
as percolating 2d surfaces. This picture fits the lattice magnetic strings
described above \cite{gorsky2}.

Furthermore, the surfaces should be topologically charged since
the D2 branes are wrapped around the $x_4$ direction.
This prediction also fits the lattice data which indicate that
the topological fermionic modes are indeed strongly
correlated with the magnetic strings.

What happens at $T=T_c$ according to the dual model?
As is mentioned above, see (\ref{geometry2}), the radius of the (Euclidean)
$\tau$-circle is now vanishing at the horizon
while the radius of the $x_4$ dimension does not depend on $u$.
Thus, wrapping around the
$x_4$ circle does not ensure  a vanishing  action any longer.
However, the D2 branes can still be dynamical degrees of freedom if they are
wrapped around the $\tau$-circle. In the geometric language wrapping around the time direction
means that the defects become time oriented.


If a surface becomes parallel to the Euclidean time, then
 the time dependence is trivial and one can concentrate
 on a 3d time slice
 of the lattice. Moreover, the intersection of the surface with the time
 slice is a line. And in the language of the quantum geometry lines represent particles
 in any number of dimensions. 
 The infinite percolating cluster of surfaces at $T>T_c$ is becoming
 an infinite percolating cluster of trajectories in 3d.

\section{Towards applications}

\subsection{Massless modes, stringy $U(1)$ symmetries}

Since we are discussing distances much larger than $\Lambda_{QCD}^{-1}$
we are in fact interested in gapless excitations. Moreover, 3d
massless scalars
are known to be a signature of superfluidity. Thus, if we find such a scalar 
in the realistic set up of the Yang-Mills theories, there are good chances
that the quark-gluon plasma contains a superfluid component.
What are the mechanisms of generating massless scalars?
First, Goldstone bosons appear in case of spontaneous symmetry breaking, or
non-vanishing vacuum expectation value
of a complex field. Massless modes can also be
related to defects with finite action, as zero modes. Both mechanisms
could be relevant to the Yang-Mills plasma.
 
 In this context, the central point
is that there exist specific stringy $U(1)$ symmetries associated with compact
dimensions. As an example, consider temperature $T=0$ and the compact
$x_4$ coordinate. Then the states of strings wrapped around this coordinate
would have a $U(1)$ charge which is nothing else but the wrapping number
$$Q~=~n_{wr}~~.$$ 
Moreover, since the $x_4$ direction is associated with the topological charge,
the wrapping number $n_{wr}$ fixes also the topological charge of the corresponding state.
Naively, the corresponding current would
 look like
\begin{equation}\label{current2}
J_{\mu}~=~q_{top}\Phi_1^{*}\partial_{\mu}\Phi_1~+~2q_{top}\Phi_2^{*}\partial_{\mu}\Phi_2~+~...,
\end{equation}
where $\Phi_{1,2}$ are wave functions of the state with $n_{wr}=1,2$
and $q_{top}$ is a unit charge. Generically, all such states are heavy and
cannot be consistently used in our model.

However, the cigar-shaped geometry in the coordinates $(u,x_4)$
suggests that the corresponding $U(1)$ symmetry can well be spontaneously broken.
Then there appears a massless Goldstone, $\phi_{\theta}$, and the effective
Lagrangian    for this field would be axion-like:
\begin{equation}
L((\phi_{\theta})~=~1/2(\partial_{\mu}\phi_{\theta})^2+f_{\theta}\partial_{\mu}\phi_{\theta} K_{\mu}+...~,
\end{equation}
where the dots involve states with masses of order $\Lambda_{QCD}$ and should be neglected in
our approximation and $K_{\mu}$ is the topological Chern-Simons current. The current 
(\ref{current2})becomes in the approximation of light states
$J_{\mu}~\sim~f_{\theta}^2\partial_{\mu}\theta +K_{\mu}$.

Another $U(1)$ symmetry, much more discussed in the literature \footnote{for an incomplete
list see \cite{kogan}.} is the topological symmetry associated with the wrapping around
the compact Euclidean type. In the deconfining phase, $T>T_c$,  the cigar-shape
geometry is in the coordinates $(u,\tau)$ and this symmetry could well be spontaneously
broken. Denote the corresponding Goldstone as $\phi_{thermal}$. Then the corresponding
$U(1)$ current would look as
\begin{equation}\label{current}
J_{i}~=~f_{th}^2\partial_i\phi_{thermal}~+...~~, i~=~1,2,3,
\end{equation} where we again do not specify the contribution of massive states.
If Eq (\ref{current}) is true then it signals superfluidity, for details
and references see \cite{verschelde},

\subsection{Massless states as vibration of branes} 

As we discussed above, the non-perturbative physics
in the Euclidean space becomes static at $T>T_c$.
At large number of colors, the first-order phase transition can be considered as a change
in the number of dimensions of non-perturbative physics \cite{nakamura}.
At $N_c=3$ the transition is somewhat smoother. 

Thus, we can consider the non-perturbative defects as living on a 3d subspace,
embedded into the 4d Euclidean space. Then the classical action associated with
the 3d brane is
\begin{equation}
S_{3d,~classical}~=~-T\cdot \int d^3x~.
\end{equation}
Let us emphasize the unconventional minus sign in right-hand side.
The common expression would of course be with the plus sign,
and $e^{-S_{classical}}$ would give the probability to excite the defect.
However, in our case the nonperturbative defects lower the energy of the vacuum.
They are always present in the 3d space. In other words, the cosmological constant
associated with the nonperturbative effects corresponds to a negative energy,
as is discussed first in connection with the gluon condensate in QCD \cite{svz}.
More formally, non-perturbative contribution to the vacuum energy is to be understood
as the difference between the energy of the true and perturbative vacuum.
This difference is negative.

In the quasi-classical approximation the action becomes
\begin{equation}\label{3d}
S_{3d,~quasi-classical}~=~-T\cdot\int d^3x \sqrt{1+(\partial_i\phi)^2}~,
\end{equation}
where $\phi$ is a massless 3d state, the physical meaning of $\phi$ being $\phi=\tau$
where $\tau$ is the
Euclidean time. The massless particle
 describes vibrations of the 3d brane in the Euclidean time direction.
Continuing (\ref{3d}) to the Minkowski space  we change the overall sign
and replace $\tau\to i\tau$ to get:
\begin{equation}\label{euclidean}
S_{3d,~Minkowski}~=~+T\cdot \int d^3x\sqrt{1-(\partial_i\phi)^2}~.
\end{equation}
Eq. (\ref{euclidean}) can serve as a starting point for applications.

\subsection{Exotic liquid}

Static physics is the same in the Euclidean and Minkowski spaces. Thus,
we could apply (\ref{euclidean}) directly to Yang-Mills plasma.
However, the applications become much richer if a relativistic generalization
is known. One can argue that the exotic liquid found in \cite{skenderis}
does provide us with such a generalization. Let us describe briefly
the results of \cite{skenderis}. One considers the action of a 4d scalar field $\phi$
\begin{equation}\label{skenderis}
S~=~T\int d^4x\sqrt{-\gamma}\sqrt{-(\partial\phi)^2}~,
\end{equation}
where $\gamma$ is the determinant of the metric tensor $\gamma_{ab} (a,b=0,1,2,3)$
defined as:
\begin{equation}
\gamma_{ab}dx^adx^b~=~-r_cd\tau^2+dx_idx^i~,
\end{equation}
where $r_c$ is arbitrary at the moment. The central point is to identify normalized
gradient of $\phi$ as the 4-velocity of an ideal liquid $u_a$:
\begin{equation}\label{velocity}
u_a~=~\partial_a \phi/\sqrt{X}~, ~~X~\equiv~-(\partial\phi)^2~\equiv~(\partial_0\phi)^2-
(\partial_i\phi)^2~.
\end{equation}
Another crucial point is that the equilibrium solution is 
$$\phi_{equilibrium}~=~t~.$$
Now, we are in position to establish a relation to our model (\ref{euclidean}).
Indeed, considering static case $\phi=\phi_{equilibrium}+\delta\phi({\bf r})$ we find that
the action (\ref{skenderis}) for the 3d brane coincides with (\ref{euclidean}).
 However, the integration measure over the the time $t$ in (\ref{skenderis})
 is far from being trivial and reflects certain dynamical assumptions,
 for the background see \cite{strominger}. Namely, the
 liquid (\ref{skenderis}) is dual to Rindler space with one extra dimension.
 Note that near the horizon $r_c~\to~0$ and the 4d space is becoming 3d space,
 as it should be on the horizon. From our perspective, the crucial point is that
 as far as the Rindler horizon approximates the (temperature) horizon of the
 model (\ref{sakai}) we can use results of \cite{skenderis} to predict the
 properties of the non-perturbative component of the Yang-Mills plasma.

\subsection{Properties of the Yang-Mills plasma}

The most unusual property of the liquid (\ref{skenderis}) is that
in equilibrium it has vanishing energy and non-vanishing pressure:
\begin{equation}\label{pressure}
(T_{ab})_{equilibrium}~=~(0,p,p,p),~~~~p=1/\sqrt{r_c} ~.
\end{equation}
The physical picture behind this observation can be readily understood in
terms
of the defects we are considering. Indeed, say the D2 magnetic vortices percolate
in 3d and in this way create pressure (which is a cosmological constant 
in the 3d language). Their time dependence, on the other hand,  is trivial.
Note that pressure $p\sim r_c^{-1/2}$ and blows up at the horizon. In reality, this
behaviour should be tempered (because approximations may fail).

As is emphasized in \cite{reviewVZ} the property (\ref{pressure})
echoes properties of the liquid living on the stretched horizon of
a black hole. Also, there are properties of the liquid which are similar to 
a superfluid. In particular, in the ideal-liquid approximation
the entropy $s=0$ while after the inclusion of dissipation effects
the ratio of the viscosity to the entropy takes \cite{skenderis} the lowest value
possible, $\eta/s=1/4\pi$. Also the 4-velocity of the liquid, see (\ref{velocity}),
is rotationless, the same as for a superfluid. 

On the lattice, one can measure separately  the contribution of magnetic strings
into the equation of state of the plasma. In particular, the contribution
of magnetic strings to the trace of energy momentum tensor 
$$T^{\mu}_{\mu}~=~\epsilon -3p$$
was measured on the lattice \cite{nakamura3}. It turned out that 
$$(T^{\mu}_{\mu})_{strings}~<~0~$$
and very 
large numerically. Qualitatively, this result is in very nice agreement
with the properties of the non-perturbative component of the plasma
we are discussing. We cannot rule out, of course, that the coincidence could be accidental.

\section{Conclusions. From confinement to superfluidity?}

 We have argued that there is ample evidence on the lattice in favor of existence
 of scalar condensates both at temperature $T=0$ where the non-perturbative physics
 is four-dimensional and at temperatures above the phase transition, $T>T_c$
 where non-perturbative physics becomes three-dimensional. The dual model (\ref{sakai})
 which is in the same universality class as the Yang-Mills theories in deep infrared
 region does suggest that there are stringy $U(1)$ symmetries associated with wrapping around compact (Euclidean) coordinates.  Moreover, the cigar shape geometry in the fifth dimension suggests spontaneous breaking of the corresponding $U(1)$ symmetry, resulting in the formation of a scalar condensate.

In case of 3d physics, that is at $T>T_c$, it turned out to be possible  
to predict properties of the Yang-Mills plasma which are both highly non-trivial
and are in no contradiction with the lattice data. At this point we used the results
of Ref. \cite{skenderis} where an exotic liquid was found (motivated by
absolutely different considerations). 

We have not discussed in any detail the
low temperature physics. However, it looks probable that the vacuum
expectation of the monopole field, observed on the lattice, signifies just
the spontaneous breaking of the stringy $U(1)$ associated with the
topological charge (or extra $x_4$-coordinate), discussed above. Moreover the vanishing of
the residue for exchanges of the corresponding Goldstone particle 
might fit the vanishing of the monopole condensate in the limit $a\to 0$,
 discussed in Sect. 2.2.

\section{Acknowledgments}\nonumber
The work of VIZ is partially supported by grants Leading Scientific
Schools NSh-6260.2010.2, RFBR -11-02-01227-à
 and Federal Special-Purpose Programme 'Cadres' of the
Russian Ministry of Science and Education.


\begin{thebibliography}{99}
\bibitem{polyakov}
A. M. Polyakov,  Phys. Lett. {\bf B59}  82 (1975);\\
A. M. Polyakov,   Nucl. Phys. {\bf B120}  (1977) 429.

\bibitem{greensite}
J. Greensite,  Prog. Part. Nucl. Phys. {\bf 51} (2003)  1.
[arXiv:hep-lat/0301023].

\bibitem{polyakov3}
 A. M. Polyakov, "Gauge fields and Strings," Harwood Academic Publishers (1987).

\bibitem{resolution}
V.I. Zakharov,  in {\it  Sense of Beauty in Physics: Miniconference in Honor of Adriano Di Giacomo on his 70th Birthday}, Pisa, Italy, 26-27 Jan 2006,
[arXiv:hep-ph/0602141].


\bibitem{theory}
V.I. Zakharov,  Braz. J. Phys. {\bf 37}   (2007) 65,
[arXiv:hep-ph/0612342];\\
V.I. Zakharov,  AIP Conf.Proc. {\bf 756} 182 (2005),
[arXiv:hep-ph/0501011].

 


\bibitem{gorsky2}
A. Gorsky,   V. Zakharov,   {Phys. Rev.}  {\bf D77}  (2008) 045017,
 arXiv:0707.1284 [hep-th].

\bibitem{zhitnitsky}
A. S. Gorsky, V. I. Zakharov, A. R. Zhitnitsky,  { Phys. Rev.}  {\bf D79}
 (2009) 106003,
 arXiv:0902.1842 [hep-ph].

\bibitem{teaney}
 D. Teaney,  Prog. Part. Nucl. Phys. {\bf 62}  (2009) 451.
 
\bibitem{monopoles}
V.G. Bornyakov et al., Phys. Lett. {\bf B537} (2002)  291,
[arXiv:hep-lat/0103032];\\
V.G. Bornyakov,   P.Yu. Boyko, M.I. Polikarpov,   V.I. Zakharov,   Nucl. Phys.
{\bf B672}  (2003) 222,
[arXiv:hep-lat/0305021].

\bibitem{chernodub2002}
M.N. Chernodub,   V.I. Zakharov,  Nucl. Phys. {\bf B669} (2003)233, [arXiv:hep-th/0211267].

\bibitem{branes}
F.V. Gubarev, A.V. Kovalenko, M.I. Polikarpov, S.N. Syritsyn,
V.I. Zakharov,  Phys. Lett. {\bf B574} (2003) 136 ,
[arXiv:hep-lat/0212003].

\bibitem{witten}
E. Witten,  Adv. Theor. Math. Phys. {\bf 2 } (1998) 505,
[arXiv:hep-th/9803131].

\bibitem{aharony}
O. Aharony,  J. Sonnenschein, Sh. Yankielowicz, Annals Phys. {\bf 322}  (2007) 1420,
[arXiv:hep-th/0604161].

 \bibitem{polyakov1}
A. M. Polyakov,  Phys. Lett. {\bf B72} (1978) 477.

\bibitem{pisarski}
A. Dumitru, Y. Guo, Y. Hidaka, Ch. P. Korthals Altes, R. D. Pisarski, 
Phys. Rev. {\bf D83} (2011) 034022.  arXiv:1011.3820 [hep-ph].



\bibitem{nakamura}
M.N. Chernodub,   A. Nakamura,   V.I. Zakharov, 
Proc. Steklov Inst. Math. {\bf 272} (2011) 75, arXiv:0904.0946 [hep-ph].


\bibitem{svz}
M.A. Shifman, A.I. Vainshtein, V.I. Zakharov,
Nucl. Phys. {\bf B147} (1979) 385.
 



\bibitem{verschelde}
H. Verschelde, and V.I. Zakharov, arXiv:1012.4821 [hep-th];
M.N. Chernodub,  H. Verschelde,  V.I. Zakharov,  Nucl. Phys. Proc. Suppl.
{\bf 207-208} (2010) 325, arXiv:1007.1879 [hep-ph].

 

\bibitem{kogan}
 B. Sathiapalan,   { Phys. Rev.}  {\bf D35}   (1988) 3277;\\
Ya. I. Kogan,   { JETP Lett.}  {\bf 45}  (1987) 709;\\
J. J. Atick, and E. Witten,  {   Nucl. Phys.}  {\bf B310}  (1988) 291;\\
 A. Adams,   X. Liu, J. McGreevy, A. Saltman, E. Silverstein, JHEP {\bf 0510}  (2005) 033,
[arXiv:hep-th/0502021];\\
 G. T. Horowitz, JHEP {\bf 0508} (2005) 091, e-Print: hep-th/0506166;\\
G. T. Horowitz,   E. Silverstein,  Phys. Rev. {\bf D73} (2006) 064016,
[arXiv:hep-th/0601032];\\
M. Kruczenski, and  A. Lawrence,   { JHEP}   {\bf 0607} (2006) 031, [arXiv:hep-th/0508148].



\bibitem{skenderis}
G. Compere, P. McFadden, K. Skenderis, M. Taylor, 
arXiv:1103.3022 [hep-th].

\bibitem{strominger}
I. Bredberg, C. Keeler, V. Lysov, A. Strominger,  JHEP  {\bf 1103} (2011) 141,
arXiv:1006.1902 [hep-th],
arXiv:1101.2451 [hep-th].

\bibitem{nakamura3}
 M.N. Chernodub, A. Nakamura, V.I. Zakharov, 
 Phys. Rev. {\bf D78}  (2008) 074021,
 arXiv:0807.5012 [hep-lat].

\bibitem{reviewVZ} H. Verschelde, and V.I. Zakharov, arXiv:1106.4154 [hep-th]
 
\end{thebibliography}
\end{document}